\documentclass[pdftex,twocolumn,epjc3]{svjour3}   

\RequirePackage[T1]{fontenc}

\smartqed 

\RequirePackage{graphicx}
\RequirePackage{mathptmx}  
\RequirePackage[numbers,sort&compress]{natbib}
\RequirePackage[colorlinks,citecolor=blue,urlcolor=blue,linkcolor=blue]{hyperref}
\usepackage{amsfonts, amsmath,amssymb}
\usepackage{mathtools}
\usepackage{xcolor}
\usepackage{verbatim}

\journalname{Eur. Phys. J. C}

 \makeatletter
 
 \newcommand{\nn}{\nonumber}

 \makeatother

\begin{document}

\title{What if gravity becomes really repulsive in the future?
}


\author{Imanol Albarran\thanksref{e1,addr1,addr2}
  \and
  Mariam Bouhmadi-L\'opez\thanksref{e2,addr3,addr4} 
  \and
  Jo\~ao Morais\thanksref{e3,addr3}
}

\thankstext{e1}{e-mail: albarran.payo@ubi.pt}
\thankstext{e2}{e-mail: mariam.bouhmadi@ehu.eus}
\thankstext{e3}{e-mail: jviegas001@ikasle.ehu.eus}

\institute{Departamento de F\'{\i}sica, Universidade da Beira Interior, Rua Marqu\^{e}s D'\'Avila e Bolama 6200-001 Covilh\~a, Portugal\label{addr1}
   \and
   Centro de Matem\'atica e Aplica\c{c}\~oes da Universidade da Beira Interior, Rua Marqu\^{e}s D'\'Avila e Bolama 6200-001 Covilh\~a, Portugal\label{addr2}
   \and
   Department of Theoretical Physics, University of the Basque Country UPV/EHU, P.O. Box 644, 48080 Bilbao, Spain\label{addr3}
   \and
   IKERBASQUE, Basque Foundation for Science, 48011, Bilbao, Spain\label{addr4}
}

\date{Received: date / Accepted: date}

\maketitle

\begin{abstract}
The current acceleration of the Universe is one of the most puzzling issues in theoretical physics nowadays. We are far from giving an answer on this letter to its true nature. Yet, with the observations we have at hand, we analyse the different patterns that the gravitational potential can show in the future. Surprisingly, gravity not only can get weaker in the near future, it can even become repulsive; or equivalently, the gravitational potential may become negative. We show this remark by using one of the simplest phenomenological model we can imagine for dark energy.
We have as well reviewed the statefinder approach of these models. For completeness, we have also showed the behaviour of the density contrast of dark matter and dark energy for these simple (yet illustrative models). Our results are displayed at present and how they evolve in the future.

\end{abstract}




\section{Introduction}

Hubble's discovery was crucial for our understanding of the Universe. He showed that the Universe was evolving and not static as it was believed at that time \cite{Hubble:1929ig}. 
His discovery was based on observing that the spectrum of far away galaxies was red-shifted which implied that those galaxies were moving away from us. He even measured the galaxies radial outward velocities and realised that it followed a rule: (i) the velocities were proportional to the distances at which the galaxies were located from us and (ii) the proportionality factor was a constant, the Hubble constant. About 70 years later, two independent teams \cite{Riess:1998cb,Perlmutter:1998hx} realised that by measuring further objects, SNeIa, the Hubble constant was not quite constant as was already expected. The issue was that the deviation from the constancy was not on the anticipated direction. It was no longer enough to invoke only matter to explain those observations. A new component had to be invoked adjectivated dark, as it interacts as far as we know only gravitationally, and named energy. This component started recently fuelling a second inflationary era of the visible Universe. Of course, all these observations, and subsequent ones, are telling us how gravity behaves at cosmological scales through the kinematic expansion of our Universe \cite{Percival:2004fs,Blake:2011rj,Beutler:2012px,Ade:2015xua,Abbott:2016ktf,Satpathy:2016tct}.

This kinematic description is linked to the dynamical expansion through the gravitational laws of Einstein theory. 
To a very good approximation, we can assume that our Universe is homogeneous and isotropic on large scales and it is filled with matter (standard and dark) and dark energy, where their relative fractional energy densities are $\Omega_{\textrm m}=0.309$ and $\Omega_{\textrm d}=0.691$, respectively, at present. In addition, the current Hubble parameter is of the order of $H_0=67.74$ km s$^{-1}$ Mpc$^{-1}$. 
We have fixed those values by using the latest Planck data \cite{Ade:2015xua} but please notice that our conclusions in this paper are unaltered by choosing other values for these physical quantities. 
In what refers to dark energy, we will assume its energy density to be evolving (or not) on time and its equation of sate (EoS) parameter, $w$, to be constant; i.e. we will be considering $w$CDM model as a natural candidate to describe our Universe. As it is well known (i) for $w<-1$ the Universe would face a big rip singularity \cite{Caldwell:1999ew,Caldwell:2003vq,Starobinsky:1999yw}, i.e., the Universe would unzip itself in a finite time from now, (ii) for $w=-1$ the Universe would be asymptotically de Sitter and finally (iii) if $w>-1$ the Universe would be asymptotically flat{ locally; i.e. the scalar curvature and the Ricci tensor would vanish for large scale factors}. As we next show this pattern is shown as well on the behaviour of the gravitational potential. 

The paper is organised as follows: in Section~\ref{background}, we review briefly the models to be considered and compare them using a cosmographic/statefinder analysis. In Section~\ref{perturbations}, we present the cosmological perturbations of the models focusing on the asymptotic behaviour of the gravitational potential. Finally, in Section~\ref{conclusion}, we conclude. In the \ref{statefinder_wCDM}, we include some formulas useful to Section~\ref{background}.




\section{Background Approach}
\label{background}

The geometry of the cosmological background is adequately given by the Friedmann-Lema\^{i}tre-Robertson-Walker line element:
\begin{align}
\label{bkgd_lineelem}		
	ds^{2} 
	= 
	-dt^2 + a^2\delta_{ij}dx^idx^j
	\,,
\end{align}
where $t$ is the cosmic time, $a(t)$ is the scale factor and $\delta_{ij}$ is the flat spatial metric. On the other hand, the matter content of the Universe can be separated in three main components: radiation, nonrelativistic matter (baryons and dark matter (DM)) and dark energy (DE). For simplicity, we model these three components using a perfect fluid description where each fluid has energy density $\rho_\mathrm{i}$ and pressure $p_\mathrm{i}=w_\mathrm{i}\rho_\mathrm{i}$. Here, i stands for radiation (r) with $w_\mathrm{r}=1/3$, for nonrelativistic matter (m) with $w_\mathrm{m}=0$, and for DE (d) with $w_\mathrm{d}=w$. The Friedmann equation for such model can be written as
\begin{align}
	\label{friedmann}
	\frac{H^2}{H_0^2} 
	= 
	\Omega_\mathrm{r,0}\left(\frac{a_0}{a}\right)^4
	+ \Omega_\mathrm{m,0}\left(\frac{a_0}{a}\right)^3
	+ \Omega_\mathrm{d,0}\left(\frac{a_0}{a}\right)^{3\left(1+w\right)}
	\,,
\end{align}
where the various $\Omega_\mathrm{i,0}:=\kappa^2\rho_\mathrm{i,0}/(3H_0^2)$ represent the present day fractional energy density of the different fluids and satisfy the constraint $1 = \Omega_\mathrm{r,0} + \Omega_\mathrm{m,0} + \Omega_\mathrm{d,0}$.
In this work, we adopt three different values for $w$: $\{-0.99,\, -1,\,-1.01\}$, in order to obtain three qualitatively different types of late-time behaviour for DE: quintessence ($w\gtrsim-1$), cosmological constant ($w=-1$) and phantom behaviour ($w\lesssim-1$).

In a cosmographic approach \cite{Visser:2004bf,Cattoen:2007sk,Capozziello:2008qc,Morais:2015ooa}, the scale factor is Taylor expanded around its present day value $a_0:=a(t_0)$ as
\begin{align}
	\label{scale factor expansion}
	\frac{a\left(t\right)}{a_0}=1+\overset{\infty}{\underset{n=1}{\sum}}\frac{A_{n}\left(t_{0}\right)}{n!}\left[H_{0}\left(t-t_{0}\right)\right]^{n}
	\,.
\end{align}
Here, $H_0$ is the present day value of the Hubble rate $H:=\dot{a}/a$, where a dot represents a derivative with respect to the cosmic time, and the cosmographic parameters $A_n$ are defined as $A_{n} := a^{\left(n\right)} / (a\,H^{n})$, $n\in \mathbb{N}$, where $a^{\left(n\right)}$ is the $n^{th}$-derivative of the scale factor with respect to the cosmic time%
\footnote{The parameters $A_{2},\, A_{3},\, A_{4},\, A_{5}$ are also known as the deceleration parameter $q=-A_{2}$, the jerk $j=A_{3}$, the snap $s=A_{4}$ and the lerk $l=A_{5}$, respectively \cite{Visser:2004bf}.}%
. 
Based on the cosmographic expansion \eqref{scale factor expansion}, the statefinder hierarchy was developed as a tool to distinguish different DE models \cite{Sahni:2002fz,Alam:2003sc,Arabsalmani:2011fz,Li:2014mua}. In fact, the statefinder parameters are defined as specific combinations of the cosmographic parameters:
\begin{align}
	\label{sn3}
	S_{3}^{\left(1\right)}= &~ A_{3}
	\,,
	\\
	\label{sn4}
	S_{4}^{\left(1\right)}= &~ A_{4}+3\left(1-A_2\right)
	\,,
	\\
	\label{sn5}
	S_{5}^{\left(1\right)}= &~ A_{5}-2\left(4-3A_2\right)\left(1-A_2\right)
	\,,
\end{align}
such that, by construction, $S_{n}^{(1)}\vert_{\Lambda\textrm{CDM}}=1$, i.e., the statefinder hierarchy defines a null diagnostic for the $\Lambda$CDM model \cite{Arabsalmani:2011fz}. It is also convenient to introduce the {statefinder} parameter $s$ defined in \cite{Sahni:2002fz} as
\begin{align}
\label{defparameter-s}
	{
	s=\frac{1-S_{3}^{\left(1\right)}}{3\left(A_2+\frac12\right)}.
	}
\end{align}

For the case of a $w$CDM model with a radiation component, such as the models considered in this paper, we present in \ref{statefinder_wCDM} the full expressions of the statefinder parameters as functions of the scale factor $a/a_0$ and the cosmological parameters $\{\Omega_\textrm{i,0},\,w\}$. In the limit $a\rightarrow+\infty$ the expressions found reduce to
\begin{align}
	\label{sn3_asymptotic}
	S_{3}^{(1)}\vert_{w\textrm{CDM}} =&~
	1 + \frac{9}{2}w\left(1+w\right)
	\,,
	\\
	\label{sn4_asymptotic}
	S_{4}^{(1)}\vert_{w\textrm{CDM}} =&~
	1 - \frac{9}{4}w\left(1+w\right)\left(7+9w\right)
	\,,
	\\
	\label{sn5_asymptotic}
	S_{5}^{(1)}\vert_{w\textrm{CDM}} =&~
	1 + \frac{9}{4}w\left(1+w\right)\left(41+87w+54w^2\right)
	\,, 
	\\
	{s\vert_{w\textrm{CDM}}} =&~{1+w}
	\,.
\end{align}
We thus find that as $w$ deviates from the nominal value $-1$ the {asymptotic values of the} statefinder parameters {$S_i^{(1)}$} run away from unity. In fact, for small deviations $\Delta w :={| w+1|}\ll1$ the statefinder parameters depend linearly on  $\Delta w$ and we find that $S_{n}^{(1)}<1$ for quintessence models and $S_{n}^{(1)}>1$ in the case of phantom behaviour.{ On the other hand, it can be shown that asymptotically $s$ vanishes for $\Lambda$CDM, and it gets negative for $w<-1$ and positive for $-1<w$. We have assumed on all our conclusions the presence of radiation no matter its tiniest contribution.}

\begin{figure}[t]
\begin{minipage}{\columnwidth}
\centering
\includegraphics[width=\columnwidth]{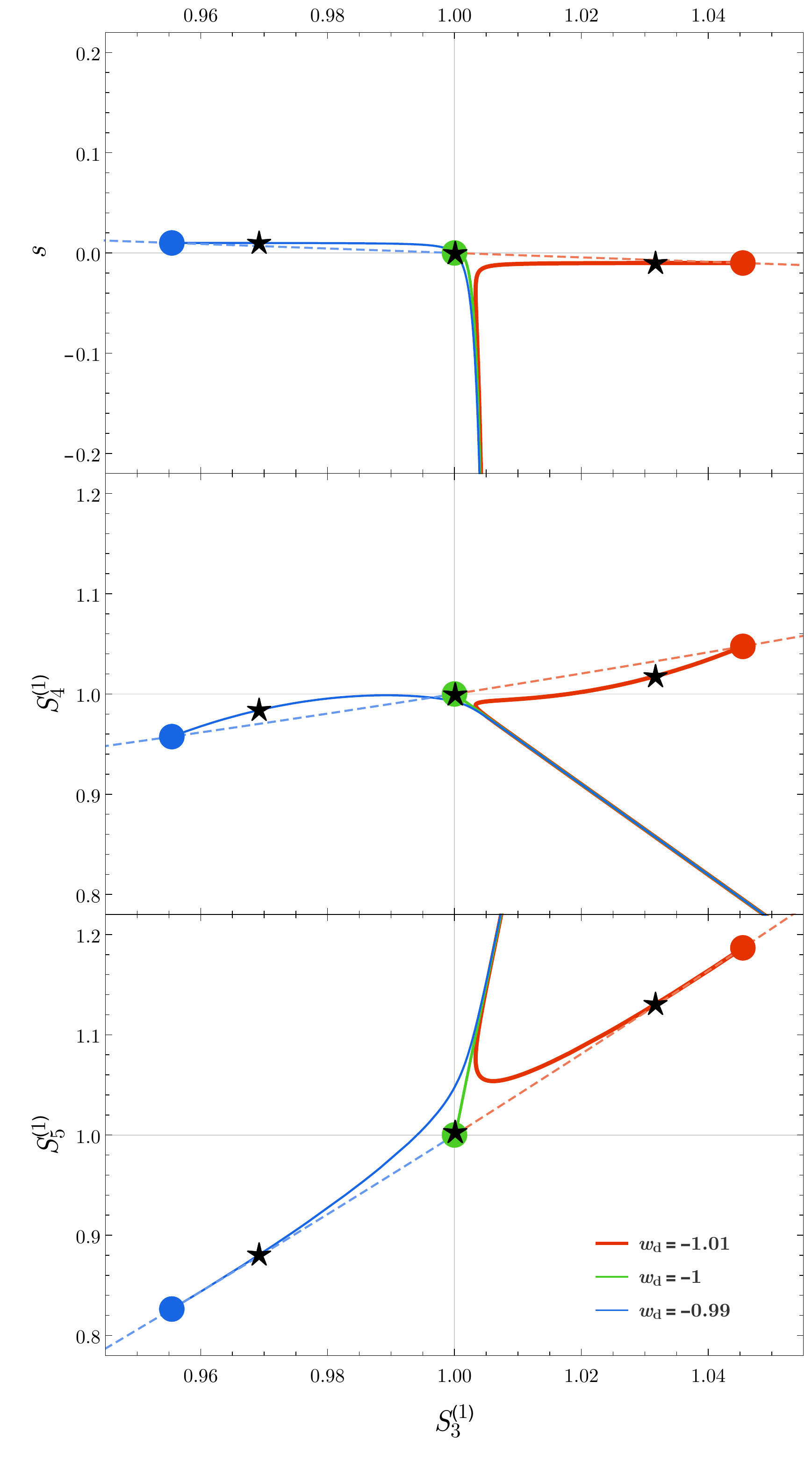}
\end{minipage}
\caption{This figure shows the trajectory of the three models considered in this work in the planes {$\{S_3^{(1)},\,s\}$},   $\{S_3^{(1)},\,S_4^{(1)}\}$ and $\{S_3^{(1)},\,S_5^{(1)}\}$ that characterise the statefinder hierarchy. The coloured points indicate the asymptotic values of the statefinder parameters as presented in eqs.~\eqref{sn3_asymptotic}--\eqref{sn5_asymptotic}. The dependence of these points on the deviation of $w$ from the $\Lambda$CDM value $-1$ is illustrated by the dashed lines. The black stars indicate the present day values of the statefinder parameters for each of the models.
}
\label{fig_S3S4S5}
\end{figure}

On Fig.~\ref{fig_S3S4S5}, we present the evolution of the statefinder hierarchy{ $\{S_3^{(1)},\,s\}$ (top panel),  $\{S_3^{(1)},\,S_4^{(1)}\}$ (middle panel)} and $\{S_3^{(1)},\,S_5^{(1)}\}$ (bottom panel) for the three models considered: $w=-0.99$ (blue), $w=-1$ (green) and $w=-1.01$ (red). When the Universe is dominated by radiation and matter the three models are indistinguishable and can be seen following the same straight line trajectory in the planes{ $\{S_3^{(1)},\,s\}$,} $\{S_3^{(1)},\,S_4^{(1)}\}$ and $\{S_3^{(1)},\,S_5^{(1)}\}$.
However, as DE starts to dominate at late-time the differences between the three models become apparent. { The trajectory $\{S_3^{(1)},\,s\}$ evolves towards the point (\,1,\,0) for the  $\Lambda$CDM model, then for a quintessence model that trajectory evolves towards the second quadrant in the plane $\{S_3^{(1)},\,s\}$\,, i.e. {$S_3^{(1)}<1$} and $0<s$, and, finally, for a phantom scenario the trajectory $\{S_3^{(1)},\,s\}$\, heads towards the fourth quadrant\,, i.e. {$1<S_3^{(1)}$} and $s<0$. For the second group of trajectories $\Big(\{S_3^{(1)},\,S_4^{(1)}\}$ and $\{S_3^{(1)},\,S_5^{(1)}\}\Big)$,  while the trajectories of the model with $w=-1$ evolve towards the point  (\,1,\,1) that characterises $\Lambda$CDM, in the quintessence model the trajectories evolve towards the third quadrant in both panels ($S_n^{(1)}<1$ for $n=3,4,5$). In contrast, for the model with phantom behaviour the trajectories  evolve towards the first quadrant in the planes $\{S_3^{(1)},\,S_4^{(1)}\}$ and $\{S_3^{(1)},\,S_5^{(1)}\}$ characterised by $S_n^{(1)}>1$ for $n=3,4,5$. Finally, by looking at Fig.~\ref{fig_S3S4S5}, it seems that the pair $\{S_3^{(1)},\,S_5^{(1)}\}$ are better to distinguish the model with {$w<-1$} from {$-1<w$}.}




\section{Cosmological Perturbations: from gravity to DM and DE}
\label{perturbations}

The gravitational potential can be described through the time-time metric component as 
\begin{align}
\label{lineelem}
	ds^{2}=a^2\left[-\left(1+2\Phi\right)d\eta^2+\left(1-2\Phi\right)\delta_{ij}dx^idx^j\right]
	\,,
\end{align}
where $\eta$ is the conformal time, $\delta_{ij}$ is the flat spatial metric and $\Phi$ the gravitational potential. For simplicity and from now on, we assume the absence of anisotropies; i.e. the spatial and temporal component of the gravitational metric are equal on absolute values at first order on the cosmological perturbations.

In order to tackle the cosmological perturbations of a perfect fluid with a negative and constant EoS some care has to be taken into account \cite{Albarran:2016mdu}. In fact, unless non-adiabatic perturbations are taken into account a blow up on the cosmological perturbations quickly appears even at scales we have already observed. Please notice that this is so even for non-phantom fluids, i.e., for $w\geq-1$. This will be our first assumption and therefore non-adiabatic perturbations will be considered. The non-adiabaticity implies the existence of two distinctive speed of sounds for the dark energy fluid: (i) its quadratic adiabatic speed of sound $c_a^2=w$ (in our case) and (ii) its effective quadratic speed of sound, $c_s^2$, whose deviation from 
$c_a^2=w$ measures the non-adiabaticity in the evolution of the fluid 
\cite{Valiviita:2008iv}. For simplicity, we will set the latter to one which fits perfectly the case of a scalar field, no matter if it is a canonical scalar field of standard or phantom nature\footnote{As long as the speed of sound $c_s^2$ is not too close to zero and \mbox{$w\simeq-1$}, {the value of $c_s^2$} will not affect so much the perturbations of dark matter. A full discussion on the effect of {the speed of sound of DE} on the perturbations of the late Universe can be found in \cite{Bean:2003fb,dePutter:2010vy,Ballesteros:2010ks}. Therefore, our choice $c_s^2=1$ is not crucial in our study, it was taken just for simplicity and because it is common to use it in codes like CAMB and CLASS though there is no fundamental reason for such a choice.}. In addition, we will solve the gravitational equations describing the cosmological perturbations at first order using the same methodology we presented in \cite{Albarran:2016mdu}.
We remind the reader that the temporal and spatial components of the conservation equation of each fluid imply \cite{Albarran:2016mdu}
\begin{align}
	\label{rad_1}
	\delta_{\mathrm{r}}'
	=&~
	4\left(\frac{k^2}{3} v_{\mathrm{r}} + \Phi'\right)
	\,,
\end{align}
\begin{align}
	\label{rad_2}
	v_{\mathrm{r}}'
	=&~
	-\left(\frac{1}{4}\delta_{\mathrm{r}} + \Phi\right)
	\,,
\end{align}
\begin{align}
	\label{mat_1}
	\delta_{\mathrm{m}}' 
	=&~
	3\left(\frac{k^2}{3} v_{\mathrm{m}} + \Phi'\right)
	\,,
\end{align}
\begin{align}
	\label{mat_2}
	v_{\mathrm{m}}' 
	=&~
	-\left(\mathcal{H}v_{\mathrm{m}}+ \Phi\right)
	\,,
\end{align}
\begin{align}
	\label{de_1}
	\delta_{\mathrm{d}}' 
	=&~
	3\left(w-1\right) \delta_{\mathrm{d}}
	\nn\\
	&~
	+ 3\left(1+w\right)\left\{\left[\frac{k^2}{3} + 3\mathcal{H}^2\left(1 - w\right) \right]v_{\mathrm{d}} + \Phi' \right\}
	\,,
\end{align}
\begin{align}
	\label{de_2}
	v_{{\mathrm{d}}}' 
	=&~
	- \left(\frac{1}{1+w}\delta_{\mathrm{d}}+\Phi\right) + 2\mathcal{H} v_{\mathrm{d}}
	\,,
\end{align}
while the $(00)$ and $(0i)$ components of the Einstein equations lead to \cite{Albarran:2016mdu}
\begin{align}
	\label{phi_1}
	\mathcal{H}\Phi'
	+ \left(\mathcal{H}^2 + \frac{k^2}{3}\right)\Phi
	 =&~
	 -\frac{1}{2}\mathcal{H}^2 \delta_{\mathrm{tot}}
	\,,\\
	\label{phi_2}
	\Phi' + \mathcal{H}\Phi 
	=&~
	-\frac{3}{2}\mathcal{H}^2\left(1+w_{\mathrm{tot}}\right)v_{\mathrm{tot}}
	\,.
\end{align}
In the previous equations, $\mathcal{H}:=a'/a$ is the conformal Hubble rate, $\delta_i$ and $v_i$ correspond to the density contrast and peculiar velocity of the fluid i, and we have decomposed all the perturbations into their Fourier modes.
The total quantities $w_{\mathrm{tot}}$, $ \delta_{\mathrm{tot}}$ and $ v_{\mathrm{tot}}$ found in \eqref{phi_1} and \eqref{phi_2} are defined through a proper averaging of the individual fluid values:
\begin{align}
	w_{\mathrm{tot}}
	=&~
	\frac{\sum_{\mathrm{i}=\mathrm{r},\mathrm{m},\mathrm{d}} \rho_\mathrm{i}\,w_\mathrm{i}}
	{\sum_{\mathrm{i}=\mathrm{r},\mathrm{m},\mathrm{d}}\rho_\mathrm{i}}
	\,,\\
	\delta_{\mathrm{tot}}
	=&~
	\frac{\sum_{\mathrm{i}=\mathrm{r},\mathrm{m},\mathrm{d}}\rho_\mathrm{i}\,\delta_\mathrm{i}}
	{\sum_{\mathrm{i}=\mathrm{r},\mathrm{m},\mathrm{d}}\rho_\mathrm{i}}
	\,,\\
	v_{\mathrm{tot}}
	=&~
	\frac{\sum_{\mathrm{i}=\mathrm{r},\mathrm{m},\mathrm{d}} \rho_\mathrm{i}\left(1+w_\mathrm{i}\right)\,v_\mathrm{i}}
	{\sum_{\mathrm{i}=\mathrm{r},\mathrm{m},\mathrm{d}}\rho_\mathrm{i}\left(1+w_\mathrm{i}\right)}
	\,.
\end{align}

In order to integrate \eqref{rad_1}--\eqref{de_2} (after assuming \eqref{phi_1} and \eqref{phi_2}) we impose the standard adiabatic initial conditions \cite{Albarran:2016mdu}
\begin{align}
	\label{initcond1}
	\frac{3}{4}\delta_\mathrm{r,ini}
	=
	\delta_\mathrm{m,ini}
	=
	\frac{\delta_\mathrm{d,ini}}{1+w}\approx\frac{3}{4}\delta_\mathrm{tot,ini}
	\,,
\end{align}
and
\begin{align}
	\label{velocityequal}
	v_\mathrm{r,ini}
	= v_\mathrm{m,ini}
	= v_\mathrm{d,ini}
	\approx v_\mathrm{tot,ini}
	\,,
\end{align}
while equations \eqref{phi_1} and \eqref{phi_2} imply
\begin{align}
	\Psi_\mathrm{ini}&\approx-\frac{1}{2} \delta_\mathrm{tot,ini}
	\,,\\
	\Psi_\mathrm{ini}&\approx-2\mathcal{H}_\mathrm{ini} v_\mathrm{tot,ini}
	\,.
\end{align}
These initial conditions are fully fixed by the Planck observational fit to single inflation \cite{Ade:2015xua}:
\begin{align}
	\Phi_\mathrm{ini} = \frac{2\pi}{3}\sqrt{2A_s}\left(\frac{k}{k_*}\right)^{n_s-1}k^{-3/2}
	\,,
\end{align}
where $A_s= 2.142\times10^{-9}$, $n_s=0.9667$ and the pivot scale is $k_*=0.05$ Mpc$^{-1}$.

\begin{figure}[t]
\begin{minipage}{\columnwidth}
\centering
\includegraphics[width=\columnwidth]{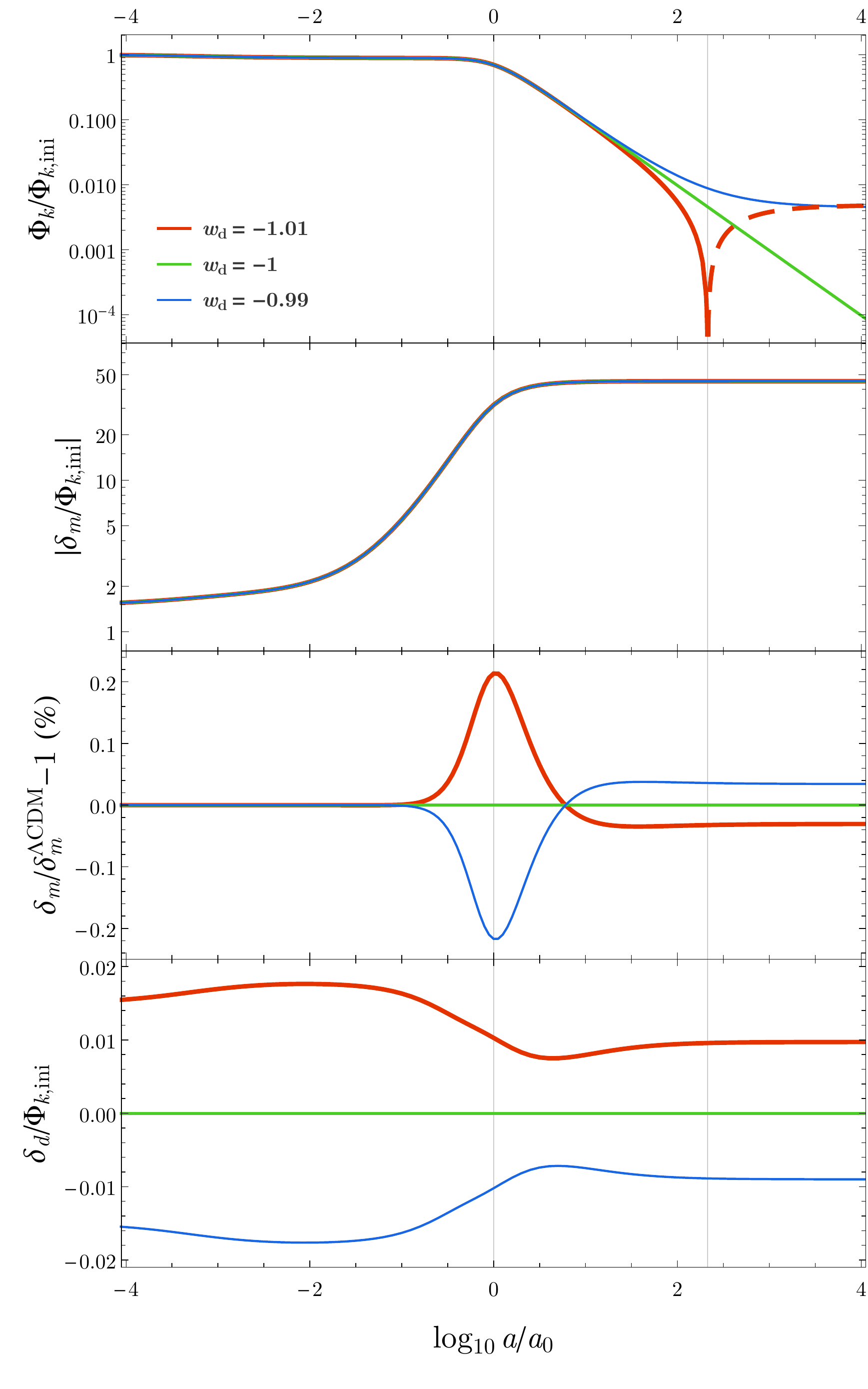}
\end{minipage}
\caption{
The evolution of the Fourier mode of the gravitational potential $\Phi_k$ (top panel), the DM perturbation $\delta_\mathrm{m}$ (middle panels) and the DE perturbation (bottom panel), from the matter era to the far future for the mode $k=10^{-3}$ Mpc$^{-1}$ and for three dark energy models: (blue) $w=-0.99$, (green) $w=-1$ and (red) $w=-1.01$. For the quintessence model (blue) the gravitational potential evolves towards a constant in the far future without changing sign, while for $\Lambda$CDM (green) $\Phi_k$ vanishes asymptotically. In the phantom model (red), $\Phi_k$ also evolves towards a constant in the far future but a change of sign occurs roughly at $\log_{10} a/a_0\simeq2.33 $, corresponding to $8.84\times10^{10}$ years in the future. A dashed line indicates negative values of $\Phi_k$.
}
\label{fig1}
\end{figure}
The behaviour of the gravitational potential and the perturbations is shown in the top panel of figure~\ref{fig1} for a given scale. We choose as an example $k=10^{-3}$ Mpc$^{-1 }$. As it must be, the gravitational potential is constant during the matter era and start decreasing as soon as dark energy \textit{goes on stage}. This behaviour is independent of the considered dark energy model. However, shortly afterwards; i.e., in our \textit{near future}, the gravitational potential will depend on the specifically chosen EoS for dark energy. In fact, (i) it will decrease until reaching a positive non-vanishing value at infinity for $w>-1$, (ii) it will vanish asymptotically for $w=-1$ and amazingly (iii) it will vanish and become negative for $w<-1$!!! This is in full agreement with the fact that close to the big rip the different structures in our Universe will be destroyed no matter their sizes or bounding energies. When could the gravitational potential vanishes and flip its sign? Of course, the answer is model and scale dependent \cite{Albarran:2016mdu}. For the model we have considered, the gravitational potential for the mode $k=10^{-3}$ Mpc$^{-1 }$ will vanish in $8.84\times10^{10}$ years from the present time or equivalently when the Universe is roughly 213 times its current size.
Furthermore, numerical results show that the smaller the scale that is considered (larger $k$) the later the gravitational potential will flip sign \cite{Albarran:2016mdu}.

In addition to the gravitational potential, we present in the second and third panels of figure~\ref{fig1} the behaviour of the density contrast of DM. We observe that the growth of the linear perturbations is very similar in all models, with differences of $\lesssim0.2\%$ with regards to $\Lambda$CDM. However, when comparing the phantom DE model with $\Lambda$CDM we find that until the present time there is an excess in the growth of the linear perturbations of DM in the phantom DE case. In the case of quintessence the opposite behaviour is observed: until the present time $\delta_\mathrm{m}$ is smaller in the quintessence case when compared with $\Lambda$CDM. 
This effect, which depends on the qualitative behaviour of DE, was first noted in \cite{Caldwell:1999ew}. Surprisingly, these deviations peak around the present time and their sign reverses in the near future. 
On the bottom panel of figure~\ref{fig1} we present the evolution of $\delta_\mathrm{DE}$ for the different models. Of course, for the $\Lambda$CDM case the perturbations remain at $0$ as the cosmological constant does not cluster. In good agreement with observations, for the quintessence and phantom DE models we find that the DE perturbations remain small, with small variations of the initial value, throughout the whole evolution of the universe.

\begin{figure}[t]
\begin{minipage}{\columnwidth}
\centering
\includegraphics[width=\columnwidth]{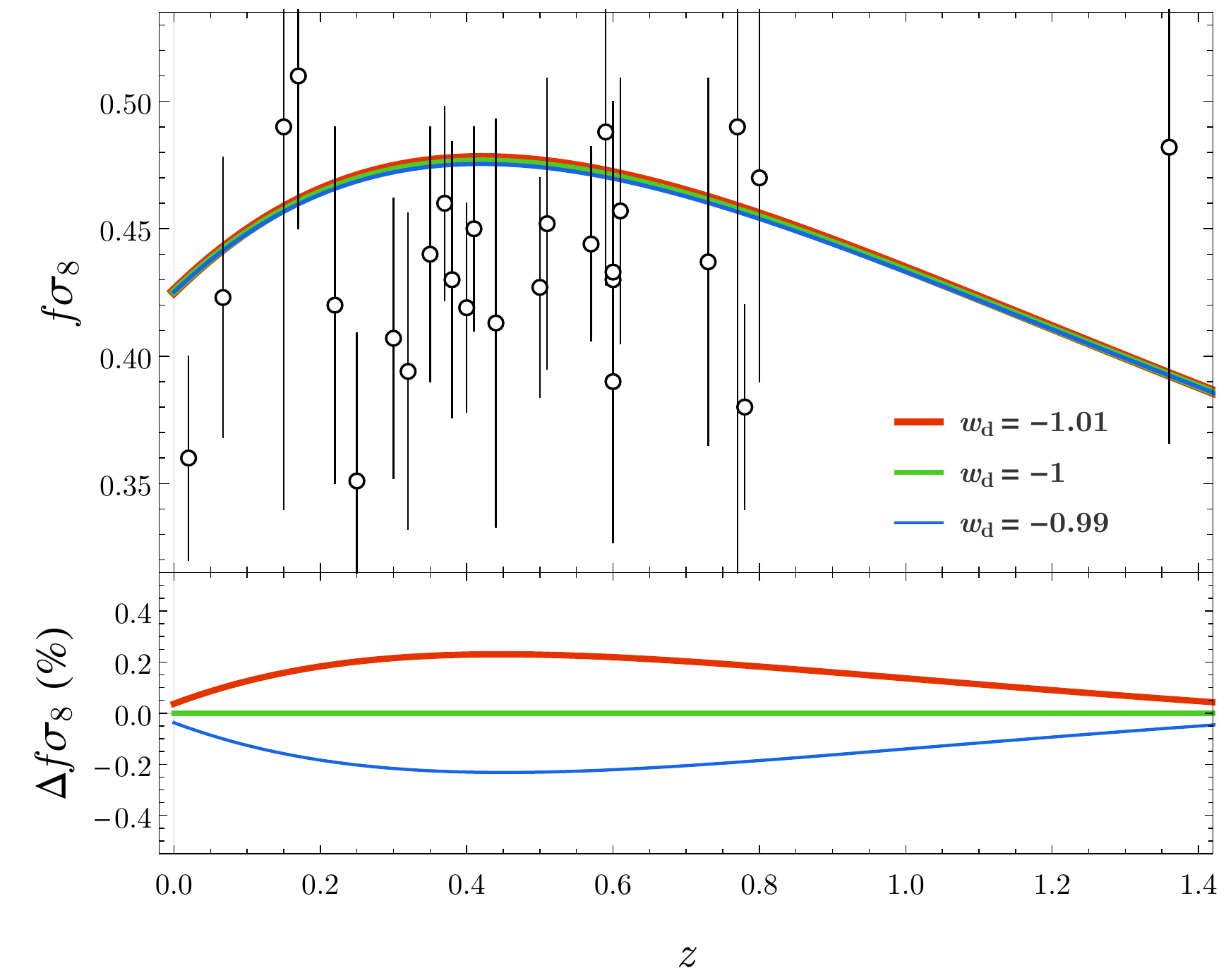}
\end{minipage}
\caption{\label{fig2}
(Top panel) evolution of $f\sigma_8$ for low red-shift $z\in(0,\,1.4)$ for three dark energy models: (blue) $w=-0.99$, (green) $w=-1$ and (red) $w=-1.01$. White circles and vertical bars indicate the available data points and corresponding error bars (cf. Table~I of \cite{Albarran:2016mdu}).
(Bottom panel) evolution of the relative differences of $f\sigma_8$ for each model with regards to $\Lambda$CDM ($w=-1$).
$\Delta f\sigma_8$ is positive in the phantom case and negative in the quintessence case.
For all the models, it was considered that $\sigma_8$ evolves linearly with $\delta_\textrm{m}$ and that $\sigma_8=0.816$ at the present time \cite{Ade:2015xua}.
}
\label{fig2}
\end{figure}

Finally, and most importantly, all these models are in full agreement with observations. 
In figure~\ref{fig2}, we show the evolution of the observable $f\sigma_8$ for the three models mentioned above. 
This combination of $f$, the relative growth of the linear matter perturbations, and $\sigma_8$, the root-mean-square mass fluctuation in spheres with radius $8$ h$^{-1}$Mpc, was proposed in \cite{Song:2008qt} as a discriminant for different models of late-time acceleration that is independent of local galaxy density bias.
On the top-panel of figure~\ref{fig2}, we contrast the $f\sigma_8$ curves of the three models with the available observational data
 (cf. Table~I of \cite{Albarran:2016mdu}). All the three curves, which are practically indistinguishable at the naked eye, are within the error bars of nearly all the points.
 On the bottom panel of figure~\ref{fig2}, we present the relative difference, $\Delta f\sigma_8$, of the results of each model with regards to $\Lambda$CDM.%
 \footnote{$\Delta f\sigma_8^\textrm{(model)} (\%):= 100[(f\sigma_8^\textrm{(model)})/(f\sigma_8^{\Lambda\textrm{CDM}})-1]$.}
Despite the small values found in terms of amplitudes, the behaviour observed suggests that the sign of $\Delta f\sigma_8$ can distinguish between a phantom (positive $\Delta f\sigma_8$) and a quintessence model (negative $\Delta f\sigma_8$). As a consequence of this difference in sign, the growth of the linear matter perturbations is stronger in a phantom scenario as opposed to $\Lambda$CDM and quintessence. This is in full agreement with the results presented in \cite{Caldwell:1999ew} where the decay of the growth suppression factor of the linear matter perturbations is found to be faster in quintessence models and slower in phantom models.



\section{Concluding remarks}
\label{conclusion}

Summarising, what we have shown is that after all gravity might behave the other way around in the future and rather than the apple falling from the tree, the apple may fly from the earth surface to the branches of the tree, if dark energy is repulsive enough, as could already be indicated by current observations\footnote{Repulsive gravity could happen as well if the effective gravitational constant  changes sign. This could happen, for example,  in scalar-tensor theories, in particular, for a non-minimally coupled scalar field \cite{Kamenshchik:2016gcy}. However, an anisotropic curvature singularity arises generically at the moment of this transition.}.

To illustrate these observations, we have considered three models where DE is characterised by a constant parameter of EoS $w$ with values $w=-0.99,-1,-1.01$. After comparing the present and future behaviour at the background level by using a statefinder approach, as illustrated in figure~\ref{fig_S3S4S5}, we have considered the cosmological perturbations of these models. We have shown that for models with $w<-1$ the gravitational potential changes its sign in the future (cf. figure~\ref{fig1}). We have as well analysed the behaviour of the DM and DE perturbations as shown for example in figure \ref{fig1}. Finally, we have proven that no matter the future behaviour of the gravitational potential depicted in figure~\ref{fig1}, the three models discussed above are in full agreement with the latest observations of $f\sigma_8$ (cf. figure~\ref{fig2}).

{ Before concluding, we would like to remind that on this work, we have considered the existence of phantom matter, however it might be possible that Nature presents rather a phantom-like behaviour as happens in brane world-models  \cite{Sahni:2002dx,Bag:2016tvc} where no big rip takes place and where the perturbations can be stable. In addition, even the presence of phantom matter might not be a problem at a cosmological quantum level where the big rip or other kind of singularities can be washed away} \cite{Dabrowski:2006dd,Kamenshchik:2007zj,BouhmadiLopez:2009pu}.

\begin{acknowledgements}
{ The work of IA was supported by a Santander-Totta fellowship ``Bolsas de Investiga\c{c}{\~a}o Faculdade de Ci{\^e}ncias (UBI)
Santander Totta''. The work of MBL is supported by the Basque Foundation of Science IKERBASQUE. {JM} is thankful to UPV/EHU for a PhD fellowship. MBL and JM acknowledge
financial support from project FIS2017-85076-P (MINECO/AEI/FEDER, UE), and Basque Government Grant No. IT956-16. This research work is supported by the grant UID/MAT/00212/2013.
This paper is based upon work from COST action CA15117 (CANTATA), supported by COST (European Cooperation in Science and Technology).}
\end{acknowledgements}

\appendix

\section{Statefinder parameters in $w$CDM}\label{app}
\label{statefinder_wCDM}

For a $w$CDM model with a radiation component the statefinder parameters defined in {eqs.~\eqref{sn3}, \eqref{sn4} , \eqref{sn5} and \eqref{defparameter-s}} read
\begin{align}
	S_{3}^{\left(1\right)}
	= &~ 
	1
	+
	\frac{
		2\Omega_\textrm{r,0}\frac{a_0}{a}
		+ \frac{9}{2} w \left(1+w\right) \Omega_\textrm{d,0} \left(\frac{a_0}{a}\right)^{3w}
	}{
		\Omega_\textrm{r,0}\frac{a_0}{a}
		+ \Omega_\textrm{m,0}
		+ \Omega_\textrm{d,0} \left(\frac{a_0}{a}\right)^{3w}
	}
	\,,
\end{align}
\begin{align}
	& S_{4}^{\left(1\right)}
	= 
	1 
	- \left[
		\Omega_\textrm{r,0}\frac{a_0}{a}
		+ \Omega_\textrm{m,0}
		+ \Omega_\textrm{d,0} \left(\frac{a_0}{a}\right)^{3w}
	\right]^{-2}
	\bigg\{
		\bigg[
			10\Omega_\textrm{r,0}\frac{a_0}{a}
		\bigg.
	\bigg.
	\nn\\
	&~~~\phantom{\times}
	\bigg.
		\bigg.
			+ 9\Omega_\textrm{m,0}
			+ \left(
				9
				+ \frac{3}{2}w\left(
					14
					+3w\left(
						7+3w
					\right)
				\right)
			\right) 
			\Omega_\textrm{d,0}\left(\frac{a_0}{a}\right)^{3w}
		\bigg]
	\bigg.
	\nn\\
	&~~
	\bigg.
		+ \frac{9}{4}w\left(1+w\right)\bigg[
			\left(7+6w\right)\Omega_\textrm{m,0}
			+ \left(7+9w\right)\Omega_\textrm{d,0} \left(\frac{a_0}{a}\right)^{3w}
		\bigg]
	\bigg.
	\nn\\
	&~~~\phantom{\times}
	\bigg.
		\times \Omega_\textrm{d,0}\left(\frac{a_0}{a}\right)^{3w}
	\bigg\}
	\,,
\end{align}
\begin{align}
	&S_{5}^{\left(1\right)}
	= 
	1 
	+ \left[
		\Omega_\textrm{r,0}\frac{a_0}{a}
		+ \Omega_\textrm{m,0}
		+ \Omega_\textrm{d,0} \left(\frac{a_0}{a}\right)^{3w}
	\right]^{-2}
	\bigg\{
		\bigg[
			76\Omega_\textrm{r,0}\frac{a_0}{a}
		\bigg.
	\bigg.
	\nn\\
	&~~~\phantom{\times}
	\bigg.
		\bigg.
			+ 60\Omega_\textrm{m,0}
			+ \bigg(
				60
				+ \frac{3}{2}w\big(
					37
					+ w\left(
						59
						+39w
						+9w^2
					\right)
				\big)
			\bigg)
		\bigg.
	\bigg.
	\nn\\
	&~~~\phantom{\times}
	\bigg.
		\bigg.
			\times
			 \Omega_\textrm{d,0}\left(\frac{a_0}{a}\right)^{3w}
		\bigg]
		\Omega_\textrm{r,0}\frac{a_0}{a} 
		+ \frac{9}{4}w\left(1+w\right)\
	\bigg.	
	\nn\\
	&~~
	\bigg.
		\times
		\bigg[
			\left(
				41
				+ 3w\left(
					17+6w
				\right)
			\right)\Omega_\textrm{m,0}
			+ \left(
				41
				+ 87w
				+54 w^2
			\right)
		\bigg.
	\bigg.
	\nn\\
	&~~~\phantom{\times}
	\bigg.
		\bigg.
			\times\Omega_\textrm{d,0} \left(\frac{a_0}{a}\right)^{3w}
		\bigg] \Omega_\textrm{d,0}\left(\frac{a_0}{a}\right)^{3w}
	\bigg\}
	\,.
\end{align}
 \begin{align}
 s=\frac{4\Omega_{\textrm{r},0}\frac{a_0}{a}+9w(1+w)\Omega_{\textrm{d},0}\left(\frac{a_0}{a}\right)^{3w}}{3\Omega_{\textrm{r},0}\frac{a_0}{a}+9w\Omega_{\textrm{d},0}\left(\frac{a_0}{a}\right)^{3w}}.\label{defstatefinder-s}
\end{align}
Due to the Friedmann constraint $1 = \Omega_\textrm{r,0} + \Omega_\textrm{m,0} + \Omega_\textrm{d,0}$ we can eliminate one of the fractional energy density parameters. It can be checked that for the $\Lambda$CDM model, where $\Omega_\mathrm{r,0}=0$ and $w=-1$ the previous expressions reduce to $S_n^{(1)}=1$ {and $s=0$}.

\end{document}